\documentclass[conference]{IEEEtran}
\usepackage{amsmath,amssymb,amsthm,epsfig,color,cite,flushend,algorithm,subcaption}
\usepackage{empheq}

\begin{document}

\title{Constructive Interference Based Secure Precoding: A New Dimension in Physical Layer Security}

\author{\IEEEauthorblockN{Muhammad R. A. Khandaker, Christos Masouros and Kai-Kit Wong}
\IEEEauthorblockA{Department of Electronic and Electrical Engineering\\
University College London\\
Gower Street, London, WC1E 7JE, United Kingdom\\
e-mail: $\{\rm m.khandaker, c.masouros, kai\text{-}kit.wong\}@ucl.ac.uk$}}

\maketitle
\begin{abstract}
Conventionally, interference and noise are treated as catastrophic elements in wireless communications. However, it has been shown recently that exploiting known interference constructively can even contribute to signal detection ability at the receiving end. This paper exploits this concept to design artificial noise (AN) beamformers constructive to the intended receiver (IR) yet keeping AN disruptive to possible eavesdroppers (Eves). The scenario considered here is a multiple-input single-output (MISO) wiretap channel with multiple eavesdroppers. Both perfect and imperfect channel information have been considered. The main objective is to improve the receive signal-to-interference and noise ratio (SINR) at IR through exploitation of AN power in an attempt to minimize the total transmit power, while confusing the Eves. Numerical simulations demonstrate that the proposed constructive AN precoding approach yields superior performance over conventional AN schemes in terms of transmit power as well as symbol error rate (SER).
\end{abstract}

\section{Introduction}
Fifth-generation ($5$G) wireless communication systems aim to achieve ultra-high spectral efficiency (SE) and orders-of-magnitude improved energy efficiency (EE). It is also expected that $5$G networks will operate in multiple tiers deploying ultra-dense small-cell base stations (BSs), e.g., heterogeneous networks (HetNets). However, a major bottleneck in ultra-dense HetNets is the cross-tier and co-tier interference. In order to harvest the full potentials of $5$G, developing sophisticated interference handling tools is a crying need at the moment. Traditional approach to dealing with the catastrophe of interference is to suppress or even kill the interference power in order to improve system performance \cite{jrnl_inter, jrnl_2way}.

However, recent developments in interference exploitation techniques have revolutionised this traditional way of dealing with known interferences \cite{const_int_lim_fac, const_int_rethink}. Constrictive interference (CI) precoding approaches suggest that interference power can even contribute to the received signal power if properly exploited \cite{const_int_lim_fac, const_int_rethink, dyn_lin_prec, const_inst_corr_rot, const_int_kn_int}. This concept imports a major breakthrough in designing wireless communication precoding when the interference is known at the transmitter. In particular downlink beamforming design can be significantly improved by symbol-level precoding of known interferences \cite{const_int_ds_cdma, const_int_miso_dl}.

The broadcast nature of wireless channels makes the communication naturally susceptible to various security threats. However, the security of wireless data transmission has traditionally been entrusted to key-based cryptographic methods at the network layer. Recently, physical-layer security (PLS) approaches have attracted a great deal of attention in the information-theoretic society since the accompanying techniques can afford an extra security layer on top of the traditional cryptographic approaches \cite{secrecy_wyner, csiszar, goel_an, qli_spatial, jrnl_secrecy, jrnl_secrecy_sinr, survey_phy_sec}. PLS exploits the channel-induced physical layer dynamics to provide information security. With appropriately designed coding and transmit precoding schemes in addition to the exploitation of any available channel state information (CSI), PLS schemes enable secret communication over a wireless medium without the aid of an encryption key.

The extent of eavesdropper's CSI available at the transmitter plays a vital role in determining the corresponding optimal transmission scheme. If full CSI of all the links is available at the transmitter, then the spatial degrees of freedom (DoF) can be fully exploited to block interception \cite{qli_spatial}. However, it is generally very unrealistic in practice. In particular, it is almost impossible to obtain perfect eavesdroppers' CSI since eavesdroppers are often unknown malicious agents. The situation can further worsen if multiple eavesdroppers cooperate in an attempt to maximize their interception through joint receive beamforming. Hence the authors in \cite{qli_spatial, jrnl_secrecy, jrnl_secrecy_sinr} considered robust secrecy beamforming design based on deterministic channel uncertainty models whereas \cite{jrnl_outage} considered probabilistic robust design.

\begin{figure*}[tb]
\centering
\begin{subfigure}{.3\textwidth}
  \centering
  \includegraphics[width=0.8\linewidth]{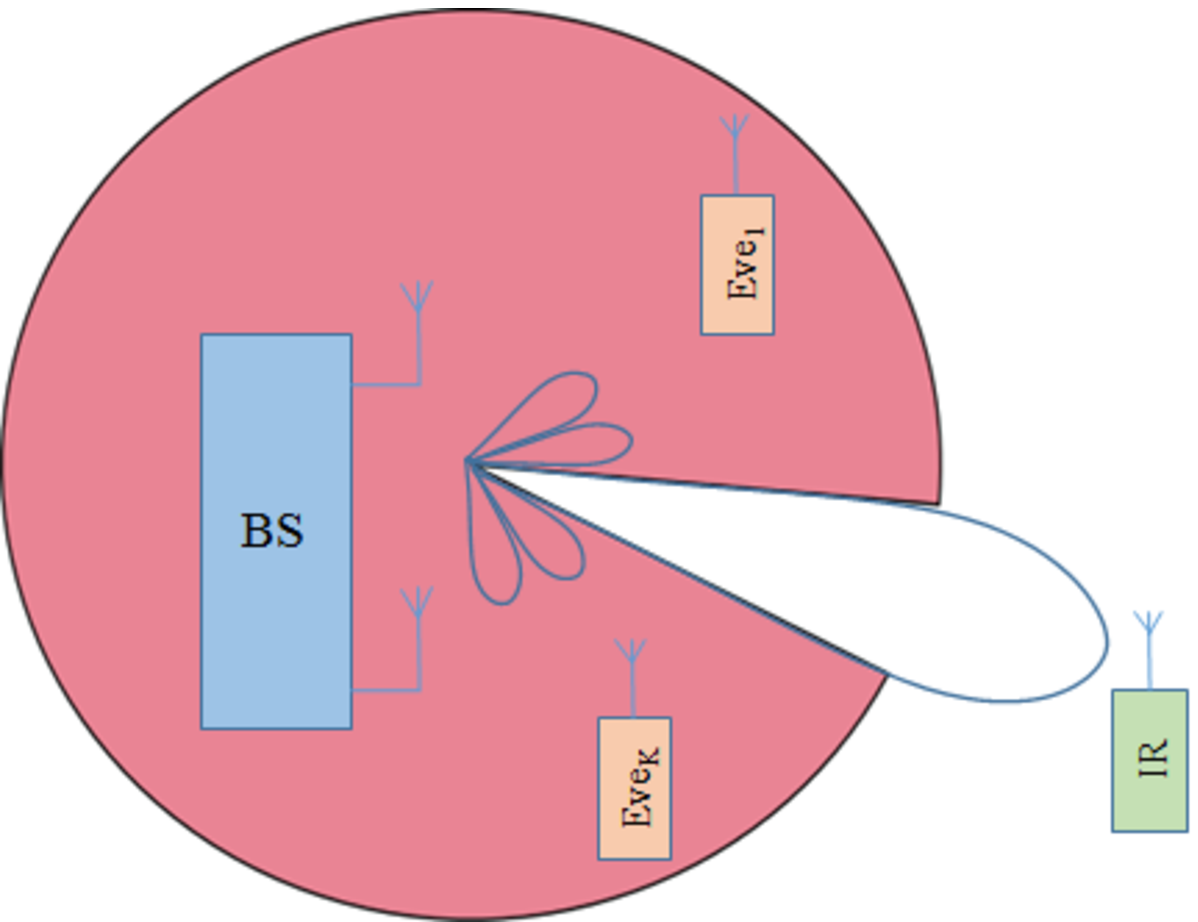}
  \caption{Conventional isotropic AN.}
  \label{sysmod_iso}
\end{subfigure}
\centering
\begin{subfigure}{.3\textwidth}
  \centering
  \includegraphics[width=.8\linewidth]{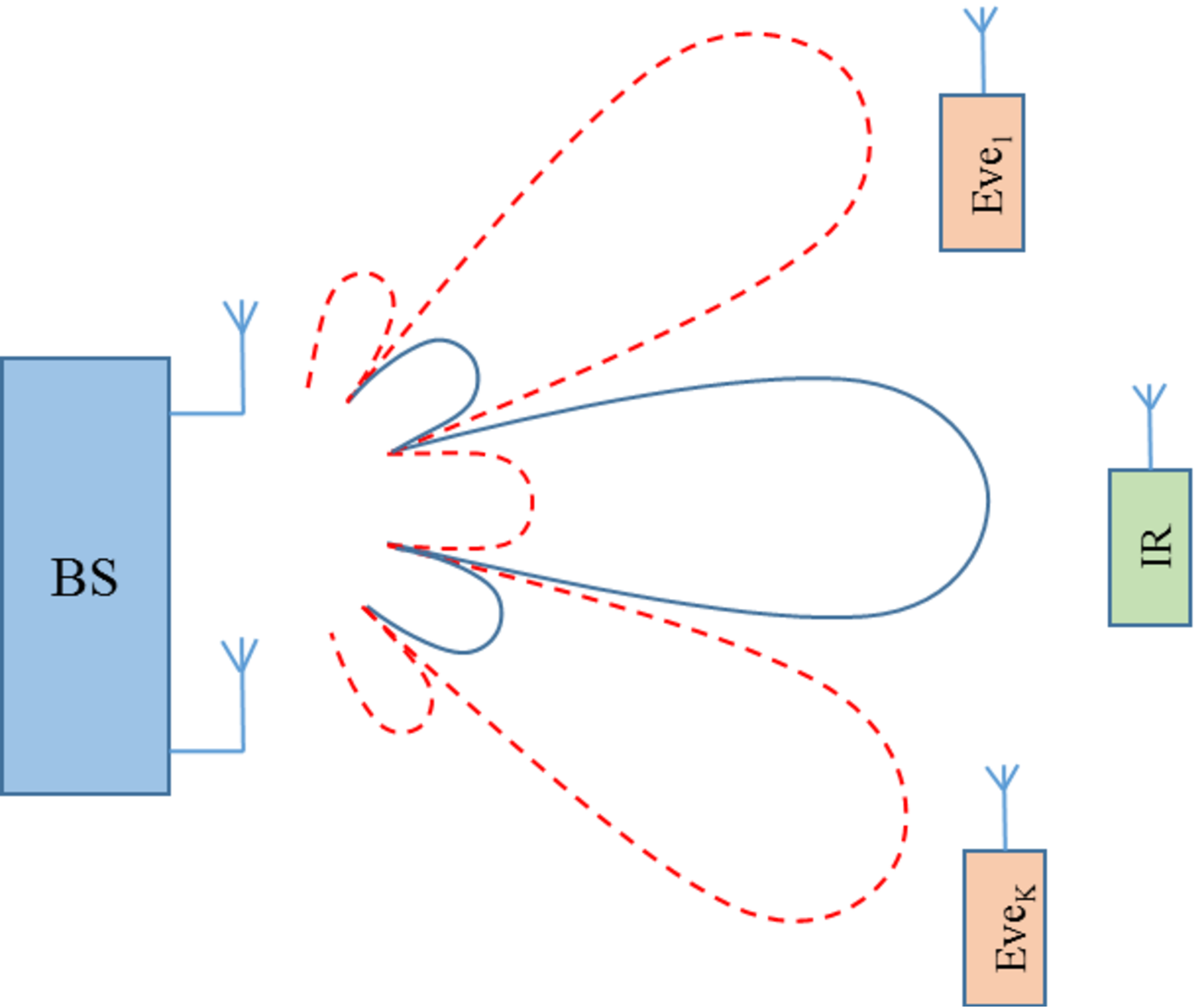}
  \caption{Conventional spatially selective AN.}
  \label{sysmod_spat}
\end{subfigure}
\begin{subfigure}{.3\textwidth}
  \centering
  \includegraphics[width=.8\linewidth]{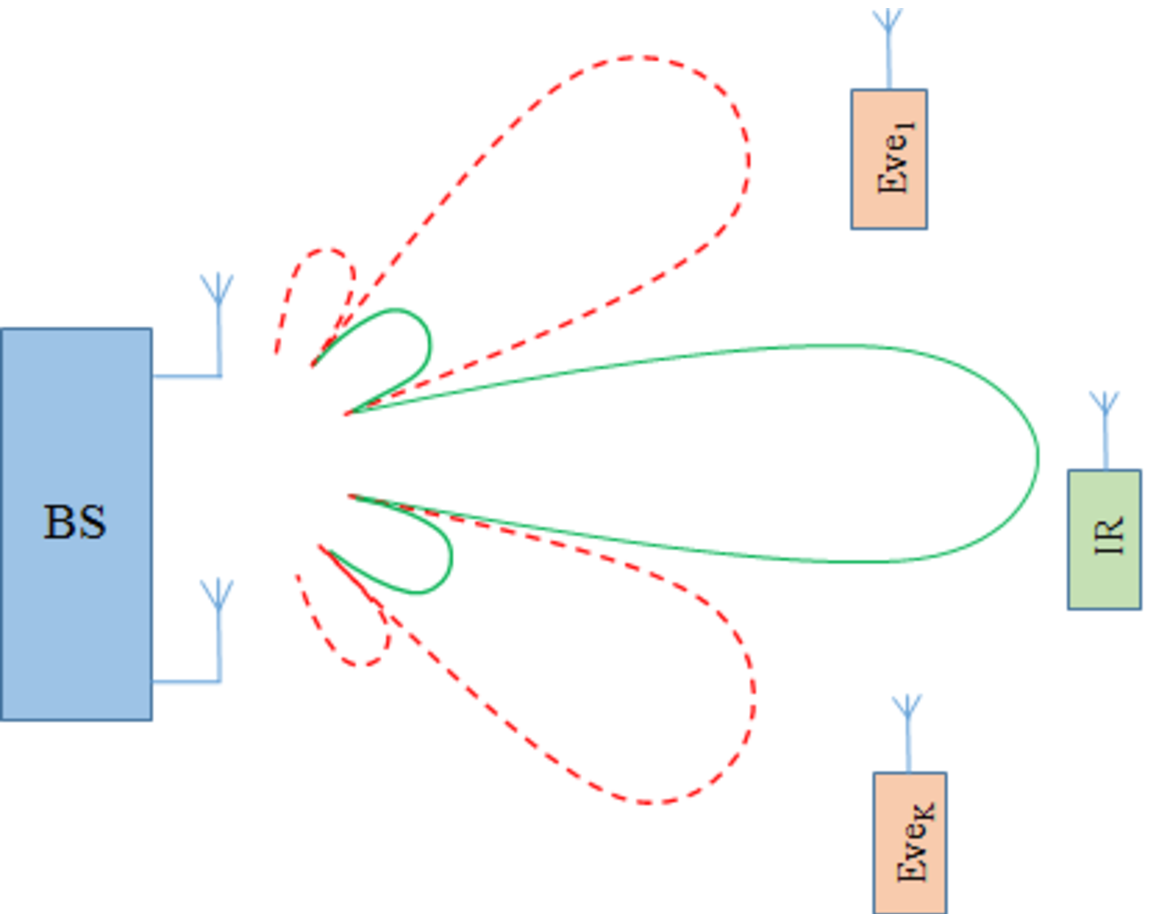}
  \caption{Constructive AN to boost received signal power.}
  \label{sysmod_const}
\end{subfigure}
\caption{Exploiting AN to boost secrecy performance.}
\label{sysmod}
\end{figure*}

To make physical-layer secrecy viable, we usually need the legitimate user's channel condition to be better than the eavesdroppers'. However, this may not always be guaranteed in practice. To alleviate the dependence on the channel conditions, recent studies showed that the spatial DoF provided by multi-antenna technology can be exploited to degrade the reception of the eavesdroppers \cite{goel_an, qli_spatial}. A possible way to do this is transmit beamforming, which concentrates the transmit signal over the direction of the legitimate user while reducing power leakage to the eavesdroppers at the same time. Apart from this, a more operational approach is to send artificially generated noise signals to interfere the eavesdroppers deliberately \cite{goel_an, qli_spatial, jrnl_secrecy, jrnl_secrecy_sinr}. Depending on the extent of eavesdroppers CSI available at the transmitter, different strategies can be applied to generate the optimal AN beams. If no eavesdroppers' CSI is available, then a popular design is the isotropic AN \cite{goel_an}, where the message is transmitted in the direction of the intended receiver's channel, and spatio-temporal AN is uniformly spread on the orthogonal subspace of the legitimate channel (see Fig.~\ref{sysmod_iso}). This scheme guarantees that the intended receiver's (IR's) reception will be free from the interference by the AN, while the ERs' reception may be degraded by the AN. On the other hand, with knowledge of the eavesdroppers' CSI to some extent, one can block the eavesdroppers' interception more efficiently by generating spatially selective AN (see Fig.~\ref{sysmod_spat}) \cite{qli_spatial, jrnl_secrecy}. More recently, an antenna array based directional modulation scheme (DM) has been studied which enhances security through adjusting the amplitude and phase of the transmit signal along a specific direction by varying the length of the reflector antennas for each symbol while scrambling the symbols in other directions\cite{dm_1st, dm_2nd, dm_sec_conf, dm_sec_jrnl}.

In this paper, we exploit the knowledge of interference available at the transmitter for improving security in wireless systems. In this context, we redesign AN signals in the form of constructive interference to the IR while keeping AN disruptive to potential Eves. We consider a multiple-input single-output (MISO) downlink system in the presence of multiple Eves as shown in Fig.~\ref{sysmod_const}. We aim at minimizing the total transmit power while boosting the received SINR at the IR as well as degrading the Eves' SINR in an attempt to keep the same below certain threshold. The benefits of constructive interference-based AN precoding scheme is twofold compared to conventional AN-based physical-layer security schemes considered in \cite{goel_an, qli_spatial, jrnl_secrecy, jrnl_secrecy_sinr}. Firstly, the constructive AN will boost the receive SINR at the IR as opposed to the conventional AN-based schemes which attempt to suppress AN signals along the direction of the IR. Secondly, to achieve a predefined level of SINR at the IR, constructive interference based precoding scheme requires lower power compared to conventional AN precoding, thus diminishing inter-user as well as inter-cell interferences. Both perfect and imperfect CSI cases have been investigated. Numerical simulations demonstrate that the proposed constructive AN precoding approach yields superior performance over conventional schemes in terms of transmit power as well as symbol error rate (SER). For clarity, the contributions are summarized below:
\begin{itemize}
\item[1)] We first consider the case when CSI is perfectly known and design two secure precoding schemes such that the AN is constructive to the IR thus reducing the required transit power for given performance and secrecy constraints.

\item[2)] We then move one step further to design a precoder such that the AN is simultentously constructive to the IR and destructive to Eves, further reducing the required transmit power  to guarantee security.
\end{itemize}
In all cases, the proposed schemes outperform the conventional AN-aided secure precoding schemes.

The rest of this paper is organized as follows. In Section~\ref{sec_sys}, the system model of a secret MISO downlink system is introduced. The conventional SINR-constrained power minimization problem is discussed in Section~\ref{sec_prob_form} for the perfect CSI case whereas in Section~\ref{sec_prob_const_int}, a constructive AN-based solution to the secrecy power minimization problem has been devised. In order to further improve secrecy performance, AN is designed as constructive for the IR and destructive for Eves in Section~\ref{sec_dest_int} considering perfect CSI. On the other hand, robust constructive-destructive AN precoding has been designed in Section~\ref{sec_rob_const_dest}. Section~\ref{sec_sim} presents the simulation results that justify the significance of the proposed algorithms under various scenarios. Concluding remarks are provided in Section~\ref{sec_con}.


\section{System Model}\label{sec_sys}
We consider a MISO downlink system where the transmitter (BS) equipped with $N_{\rm T}$ transmit antennas intend to transmit a secret message to the IR in the presence of $K$ possible eavesdroppers. The IR and the Eves are all equipped with a single antenna. In order to confuse the Eves, the BS injects AN signals into the secret message in an attempt to reduce the receive SINRs at the Eves. Thus the received signal at the IR and those at the Eves are given, respectively, by $y_{\rm d}$ and $y_{{\rm e},k}$:
\begin{align}
y_{\rm d} & = {\bf h}_{\rm d}^T{\bf x} + n_{\rm d},\label{yd}\\
y_{{\rm e},k} & = {\bf h}_{{\rm e},k}^T{\bf x} + n_{{\rm e},k}, ~\mbox{for }k=1,\dots, K,
\end{align}
where ${\bf h}_{\rm d}$ and ${\bf h}_{{\rm e},k}$ are the complex channel vectors between the BS and the IR and between the BS and the $k$th Eve, respectively, $n_{\rm d}\sim\mathcal{CN}(0,\sigma_{\rm d}^2)$ and $n_{{\rm e},k}\sim\mathcal{CN}(0,\sigma_{\rm e}^2)$ are the additive Gaussian noises at the IR and the $k$th Eve, respectively. The BS chooses ${\bf x}$ as the sum of information beamforming vector ${\bf b}_{\rm d}s_{\rm d}$ and the AN vector ${\bf b}_{\rm n} \triangleq \sum_{i=1}^N{\bf b}_{{\rm n},i}s_{{\rm n},i}$ such that the baseband transmit signal vector is
\begin{equation}
{\bf x} = {\bf b}_{\rm d}s_{\rm d} + \sum_{i=1}^N{\bf b}_{{\rm n},i}s_{{\rm n},i},
\end{equation}
where $s_{\rm d}\sim\mathcal{CN}(0,1)$ is the confidential information-bearing symbol for the IR and $s_{{\rm n},i}\sim\mathcal{CN}(0,1), \forall i,$ are the AN symbols in which $N$ denotes the number of AN symbols.


Accordingly, the received SINR at the IR is given by
\begin{equation}\label{sinr_d}
\gamma_{\rm d} = \frac{\left|{\bf h}_{\rm d}^T{\bf b}_{\rm d}\right|^2}{\sum_{i=1}^N\left|{\bf h}_{\rm d}^T{\bf b}_{{\rm n},i}\right|^2 + \sigma_{\rm d}^2},
\end{equation}
and that at the $k$th Eve is given by
\begin{equation}\label{sinr_e}
\gamma_{{\rm e},k} = \frac{\left|{\bf h}_{{\rm e},k}^T{\bf b}_{\rm d}\right|^2}{\sum_{i=1}^N\left|{\bf h}_{{\rm e},k}^T{\bf b}_{{\rm n},i}\right|^2 + \sigma_{\rm e}^2}, \forall k.
\end{equation}
The transmit signal ${\bf x}$ can also be expressed as
\begin{equation}
{\bf x} = {\bf b}_{\rm d}s_{\rm d} + \sum_{i=1}^N{\bf b}_{{\rm n},i}e^{j\left(\phi_{{\rm n},i}-\phi_{\rm d}\right)}s_{\rm d},
\end{equation}
where $s_{\rm d} = de^{j\phi_{\rm d}}$. Assuming constant envelop $d=1$, the instantaneous transmit power is given by
\begin{equation}\label{Pt}
P_{\rm T} = \left\|{\bf b}_{\rm d} + \sum_{i=1}^N{\bf b}_{{\rm n},i}e^{j\left(\phi_{{\rm n},i}-\phi_{\rm d}\right)} \right\|^2.
\end{equation}

\section{Problem Formulation}\label{sec_prob_form}
We consider a power minimization problem for secure transmission of information to the IR. Note that by minimizing the power we actually decrease the SINR at Eves while maintaining the QoS of signal reception at the desired destination. Exploitation of the available knowledge of the AN will boost the receive SINR at the IR.
In order to satisfy the secrecy requirements, conventional power minimization problem is formulated as


\begin{subequations}\label{minP_conv}
\begin{eqnarray}
{\bf{P0:}} ~~ \min_{{\bf b}_{\rm d},\{{\bf b}_{{\rm n},i}\}} \!\!\!& &\!\!\! \left\|{\bf b}_{\rm d}\right\|^2 + \sum_{i=1}^N\left\|{\bf b}_{{\rm n},i}\right\|^2\label{minP_conv_o}\\
{\rm s.t.} \!\!\!& &\!\!\! \frac{\left|{\bf h}_{\rm d}^T{\bf b}_{\rm d}\right|^2}{\sum_{i=1}^N\left|{\bf h}_{\rm d}^T{\bf b}_{{\rm n},i}\right|^2 + \sigma_{\rm d}^2}\ge \Gamma_{\rm d},\label{minP_conv_c1}\\
\!\!\!& &\!\!\! \frac{\left|{\bf h}_{{\rm e},k}^T{\bf b}_{\rm d}\right|^2}{\sum_{i=1}^N\left|{\bf h}_{{\rm e},k}^T{\bf b}_{{\rm n},i}\right|^2 + \sigma_{\rm e}^2} \le \Gamma_{{\rm e},k}, \forall k.\label{minP_conv_c2}
\end{eqnarray}
\end{subequations}
The power minimization problem has been solved considering various system configurations \cite{jrnl_secrecy, qli_qos}. One conventional approach is to reformulate problem \eqref{minP_conv} as the following SDP
\begin{subequations}\label{minP_conv2}
\begin{eqnarray}
\min_{{\bf W}_{\rm d},\{{\bf W}_{{\rm n},i}\}} \!\!\!& &\!\!\! {\rm Tr}({\bf W}_{\rm d}) + \sum_{i=1}^N{\rm Tr}({\bf W}_{{\rm n},i}) \label{minP_conv2_o}\\
{\rm s.t.} \!\!\!& &\!\!\! {\frac{1}{\Gamma_{\rm d}}} {\rm Tr}({\bf W}_{\rm d} {\bf R}_{\rm d})- \sum_{i=1}^N{\rm Tr}({\bf R}_{\rm d} {\bf W}_{{\rm n},i}) \geq \sigma_{\rm d}^{2},\label{minP_conv2_c1}\\
\!\!\!& &\!\!\! \!\! {\frac{1}{\Gamma_{{e},k}}} {\rm Tr}({\bf W}_{\rm d} {\bf R}_{{\rm e},k}) - \!\! \sum_{i=1}^N{\rm Tr}({\bf R}_{{\rm e},k} {\bf W}_{{\rm n},i}) \!\le \! \sigma_{\rm e}^{2}, \forall k,\nonumber\\ \label{minP_conv2_c2}\\
\!\!\!& &\!\!\! {\bf W}_{\rm d} \succeq {\bf 0}, ~~ {\bf W}_{{\rm n},i} \succeq {\bf 0}, \forall i, ~~ {\rm rank}({\bf W}_{\rm d})=1, \label{minP_conv2_c3}
\end{eqnarray}
\end{subequations}
where ${\bf W}_{\rm d} \triangleq {\bf b}_{\rm d}{\bf b}_{\rm d}^H$ and ${\bf W}_{{\rm n},i} \triangleq {\bf b}_{{\rm n},i}{\bf b}_{{\rm n},i}^H, \forall i$. Conventionally, the non-convex rank constraint is dropped so that the relaxed problem can be solved using existing solvers \cite{cvx}. Interestingly, it has been proved in \cite{jrnl_secrecy, qli_qos} that for a practically representative class of scenarios, the original problem can be solved optimally. Although the solutions proposed in \cite{jrnl_secrecy, qli_qos} are optimal from stochastic viewpoint, the hidden power in the AN signals has been treated as harmful for the desired information, and hence, either nullified or suppressed. In the following section, we endeavour to develop precoding schemes exploiting the AN power constructively for the desired signal at the IR.

\begin{figure*}[tb]
\centering
\begin{subfigure}{.45\textwidth}
  \centering
  \includegraphics[width=\linewidth]{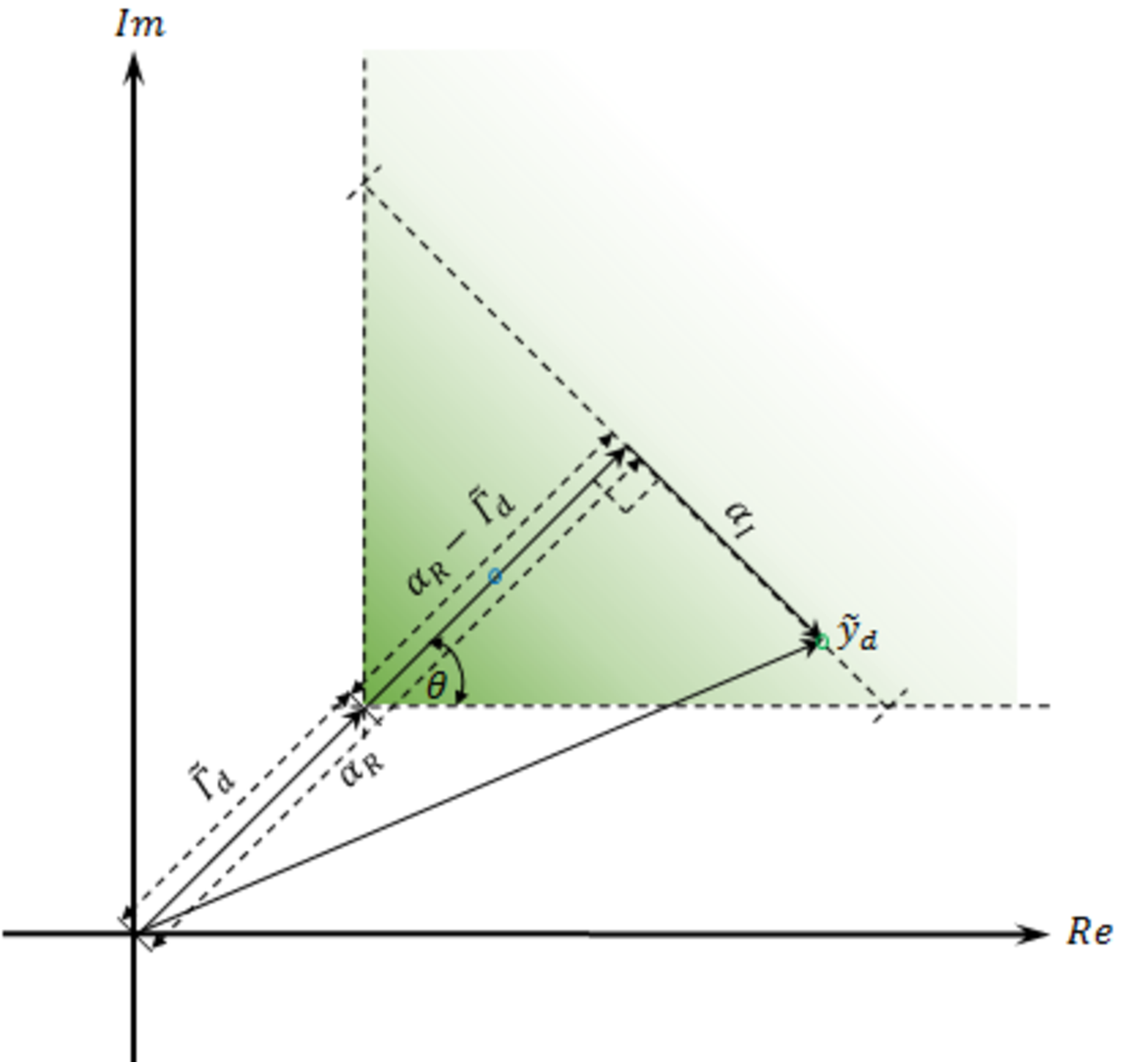}
  \caption{Constructive AN design for the legitimate receiver.}
  \label{const_bob}
\end{subfigure}
\begin{subfigure}{.45\textwidth}
  \centering
  \includegraphics[width=\linewidth]{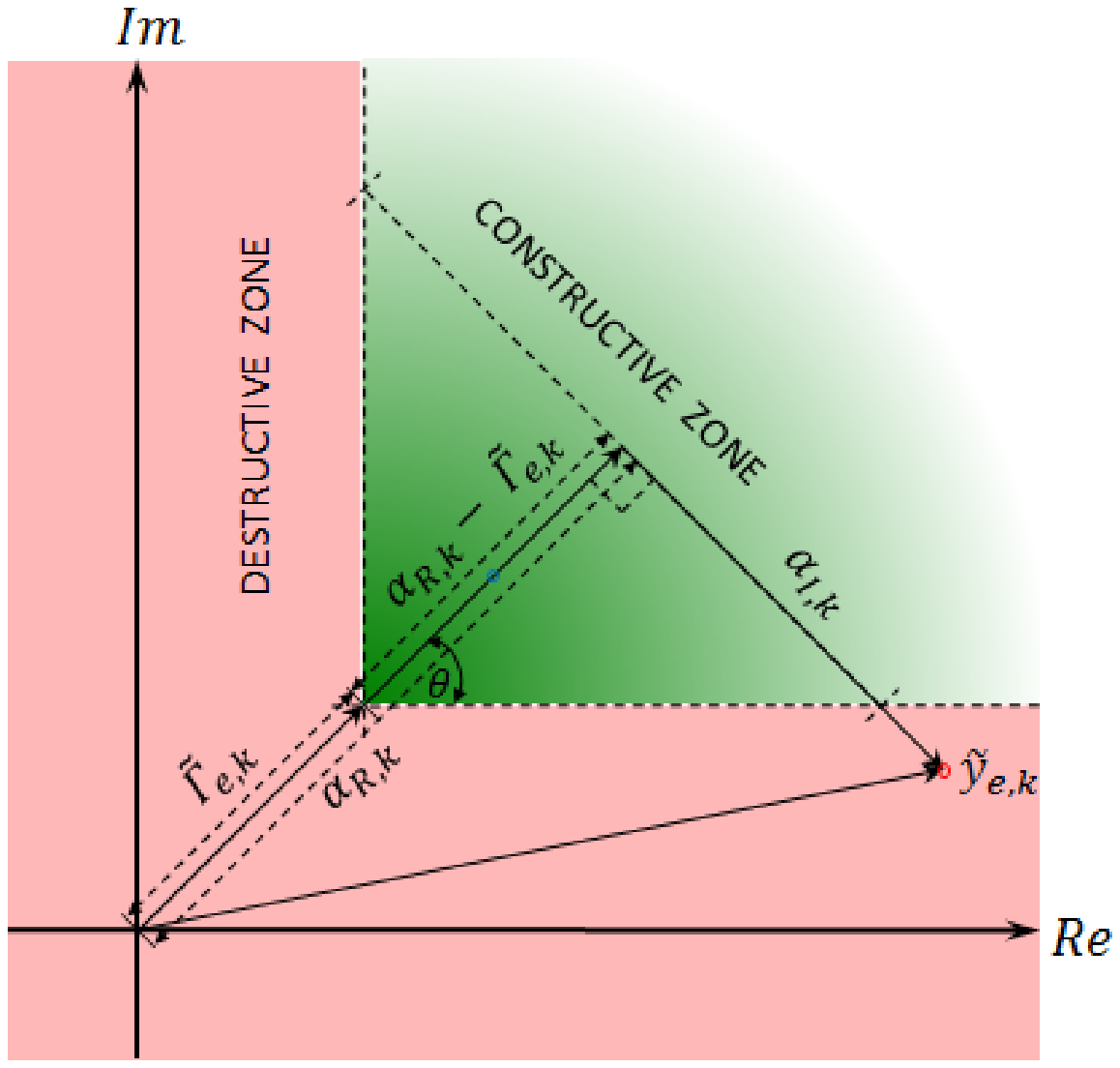}
  \caption{Destructive AN design for the eavesdropper.}
  \label{dest_eve}
\end{subfigure}
\caption{Exploiting constructive and destructive AN for QPSK symbols.}
\label{const_dest}
\end{figure*}

\section{Constructive AN-based Secure Precoding}\label{sec_prob_const_int}
In this section, we take the above approach one step forward, by actively exploiting interference (AN in this case) constructively for the IR to reduce the required power for a given SNR threshold, while guaranteeing the secrecy constraint for the Eves. We do this by optimizing the transmitted signal part ($\bf x$ in \eqref{yd}), which comprises of the desired symbol and the AN symbols. The theory and characterization criteria for constructive interference have been expensively studied in \cite{const_int_lim_fac, const_int_rethink, dyn_lin_prec, const_inst_corr_rot, const_int_kn_int, const_int_ds_cdma}. To avoid repetition, we refer the reader to the above works for the details, while here we employ this consept directly to design our new optimization problems. The AN signal will be constructive to the received signal at the IR if that moves the receives symbols away from the decision thresholds of the constellation (e.g. real and imaginary axes for QPSK symbols in Fig.~\ref{const_bob}). Hence we intend to keep the angle of that part aligned with the angle of the corresponding desired symbol $s_{\rm d}$ by appropriately designing the transmit beamforming vectors. We can do so by pushing the decision symbols towards the constructive regions of the modulation constellation, denoted by the green shaded areas (cf. Fig.~\ref{const_bob}).

For constructive precoding, the AN signals received at the IR are not suppressed or nullified in contrast to the conventional use of AN \cite{jrnl_secrecy, qli_spatial}, rather optimized instantaneously such that it contributes to the received signal power. If the AN signals can be aligned with the data symbols $s_{\rm d}$ by properly designing the beamforming precoding vectors ${\bf b}_{{\rm n},i}, \forall i$, then all AN symbols will contribute constructively. Accordingly, it can be shown that the receive SINR \eqref{sinr_d} at the IR can be rewritten as \cite{const_inst_corr_rot, const_int_kn_int}
\begin{equation}\label{sinrd_const}
\gamma_{\rm d} = \frac{\left|{\bf h}_{\rm d}^T{\bf b}_{\rm d}s_{\rm d} + {\bf h}_{\rm d}^T\sum_{i=1}^N{\bf b}_{{\rm n},i}s_{{\rm n},i}\right|^2} {\sigma_{\rm d}^2}.
\end{equation}
Note that the receive SINR at the IR has actually become SNR after constructive AN precoding. However, the SINR at the $k$th Eve remains the same as in \eqref{sinr_e} since no AN signal has been made constructive to the Eves.

Thus exploiting AN power constructively, the SINR constraint \eqref{minP_conv_c1} can be reformulated as the following system of constraints
\begin{subequations}\label{sinr_const1}
\begin{align}
\angle\left({\bf h}_{\rm d}^T{\bf b}_{{\rm d}}s_{{\rm d}} + \sum_{i=1}^N{\bf h}_{\rm d}^T{\bf b}_{{\rm n},i}s_{{\rm n},i}\right) &= \angle \left(s_{\rm d}\right)\label{sinr_const11}\\
\frac{\Re \left\{{\bf h}_{\rm d}^T\left({\bf b}_{\rm d} + \sum_{i=1}^N{\bf b}_{{\rm n},i}e^{j\left(\phi_{{\rm n},i}-\phi_{\rm d}\right)}\right)\right\}^2} {\sigma_{\rm d}^2} & \ge \Gamma_{\rm d},\label{sinr_const12}
\end{align}
\end{subequations}
where $\Re\{x\}$ indicates the real part of the complex number $x$ and $\angle x$ denotes the corresponding angle. Note that the phases of the AN signals in \eqref{sinr_const12} has been shifted by the phase of the desired symbol $s_{\rm d}$. The constraint \eqref{sinr_const11} imposes that the AN fully aligns with the phase of the symbol of interest $s_{\rm d}$ at the IR, whereas the constraint \eqref{sinr_const12} guarantees that the constructively precoded AN signals can adequately satisfy the SINR requirement at the IR.

Essentially, the angular constraint \eqref{sinr_const11} is a very strict constraint. But exploiting the concept of constructive interference, we can actually relax this constraint without losing any optimality which results in a larger feasible region. Let us denote ${\tilde y}_{\rm d} \triangleq {\bf h}_{\rm d}^T\left({\bf b}_{\rm d} + \sum_{i=1}^N{\bf b}_{{\rm n},i}e^{j\left(\phi_{{\rm n},i}-\phi_{\rm d}\right)}\right)$ as the received signal ignoring the AWGN at the IR, with constructive AN injected, and $\alpha_{\rm R}$ and $\alpha_{\rm I}$ as the abscissa and the ordinate of the phase-adjusted signal ${\tilde y}_{\rm d}$, respectively. Applying basic geometric principles to Fig.~\ref{const_bob}, the constraints in \eqref{sinr_const1} can be equivalently represented as
\begin{subequations}\label{sinr_const2}
\begin{align}
\Im \left\{{\bf h}_{\rm d}^T\left({\bf b}_{\rm d} + \sum_{i=1}^N{\bf b}_{{\rm n},i}e^{j\left(\phi_{{\rm n},i}-\phi_{\rm d}\right)}\right)\right\} & = 0\label{sinr_const21}\\
\Re \left\{{\bf h}_{\rm d}^T\left({\bf b}_{\rm d} + \sum_{i=1}^N{\bf b}_{{\rm n},i}e^{j\left(\phi_{{\rm n},i}-\phi_{\rm d}\right)}\right)\right\} & \ge \sigma_{\rm d}\sqrt{\Gamma_{\rm d}},\label{sinr_const22}
\end{align}
\end{subequations}
where $\Im\{x\}$ indicates the imaginary part of the complex number $x$. However, it can be observed from Fig.~\ref{const_bob} that the AN contaminated received signal ${\tilde y}_{\rm d}$ does not necessarily need to strictly align the angle of the desired signal. That is, ${\tilde y}_{\rm d}$ lays on the constructive zone of the desired symbol $s_{\rm d}$ as long as the following condition is satisfied
\begin{equation}
-\theta \le \phi_{\rm d} \le \theta, \quad \text{i.e.,} \quad \frac{\left| \alpha_{\rm I} \right|}{\alpha_{\rm R}-\tilde{\Gamma}_{\rm d}} \le \tan\theta,
\end{equation}
where $\tilde{\Gamma}_{\rm d} \triangleq \sigma_{\rm d}\sqrt{\Gamma_{\rm d}}$ and $\theta = \pi/M$, $M$ is the constellation size. Thus the strict angle constraint \eqref{sinr_const21} can be relaxed as
\begin{multline}\label{ang_const}
\left|\Im \left\{{\bf h}_{\rm d}^T\left({\bf b}_{\rm d} + \sum_{i=1}^N{\bf b}_{{\rm n},i}e^{j\left(\phi_{{\rm n},i}-\phi_{\rm d}\right)}\right)\right\}\right|\\
\le \left(\Re \left\{{\bf h}_{\rm d}^T\left({\bf b}_{\rm d} + \sum_{i=1}^N{\bf b}_{{\rm n},i}e^{j\left(\phi_{{\rm n},i}-\phi_{\rm d}\right)}\right)\right\} - \sigma_{\rm d}\sqrt{\Gamma_{\rm d}}\right)\\
\times\tan\theta,
\end{multline}
Interestingly, the QoS constraint \eqref{sinr_const22} is embedded in \eqref{ang_const}. Hence we do not need to explicitly include it in the constructive interference precoding optimization problem. Thus the plain constructive interference based secure transmit precoding optimization problem can be formulated as
\begin{subequations}\label{minP_const}
\begin{eqnarray}
{\bf{P1:}} ~~ \min_{{\bf b}_{\rm d},\{{\bf b}_{{\rm n},i}\}} \!\!\!& &\!\!\! \left\|{\bf b}_{\rm d} + \sum_{i=1}^N{\bf b}_{{\rm n},i}e^{j\left(\phi_{{\rm n},i}-\phi_{\rm d}\right)} \right\|^2\label{minP_const_o}\\
{\rm s.t.} \!\!\!& &\!\!\! \left|\Im \left\{{\bf h}_{\rm d}^T\left({\bf b}_{\rm d} + \sum_{i=1}^N{\bf b}_{{\rm n},i}e^{j\left(\phi_{{\rm n},i}-\phi_{\rm d}\right)}\right)\right\}\right| \nonumber\\
\!\!\!& &\!\!\! \le \left(\Re \left\{{\bf h}_{\rm d}^T\left({\bf b}_{\rm d} + \sum_{i=1}^N{\bf b}_{{\rm n},i}e^{j\left(\phi_{{\rm n},i}-\phi_{\rm d}\right)}\right)\right\}\right. \nonumber\\
\!\!\!& &\!\!\! \qquad\qquad\qquad \left. - \sigma_{\rm d}\sqrt{\Gamma_{\rm d}}\right)\tan\theta,\label{minP_const_c1}\\
\!\!\!& &\!\!\! \frac{\left|{\bf h}_{{\rm e},k}^T{\bf b}_{\rm d}\right|^2}{\sum_{i=1}^N\left|{\bf h}_{{\rm e},k}^T{\bf b}_{{\rm n},i}\right|^2 + \sigma_{\rm e}^2} \le \Gamma_{{\rm e},k}, \forall k.\label{minP_const_c2}
\end{eqnarray}
\end{subequations}
Note that problem \eqref{minP_const} adopts the instantaneous transmit power (including data symbols) as the objective to minimize, as opposed to the average transmit power in conventional optimization framework \eqref{minP_conv}. Manipulating the constraint \eqref{minP_const_c2}, the problem \eqref{minP_const_c2} can be reformulated as a standard SOCP, which can be optimally solved using optimization toolboxes, e.g., CVX \cite{cvx}. Note that the relaxed angular constraint \eqref{minP_const_c1} allows a larger feasibility region (entire green zone in Fig.~\ref{const_bob}), which results in a lower minimum transmit power as we will observe in Section~\ref{sec_sim}.

\section{Destructive Interference Based AN Precoding}\label{sec_dest_int}
In this section, our attempt is to further improve the secrecy performance utilizing the concept of destructive interference for the eavesdroppers. The concept is that, we will design the AN beamformers such that the AN signal is constructive to the IR while destructive to the Eves. As long as some knowledge of the Eves' channels is available at the transmitter, one can do so by pushing the received signal at the IR towards the decision thresholds (green zone in Fig.~\ref{const_bob}) while pushing the received signal at the Eves away from the decision thresholds (red zone in Fig.~\ref{dest_eve}). This makes correct detection more challenging for the Eves by reducing the receive SINR. The benefit is that given secrecy thresholds can be guaranteed with lower transmit power. More importantly, it will be shown in the following optimization schemes that the secrecy constraints are guaranteed on a symbol-by-symbol basis, rather than the conventional statistical guarantees.

By denoting $\alpha_{{\rm R},k}$ and $\alpha_{{\rm I},k}$ as the real and imaginary parts of ${\tilde y}_{{\rm e},k} \triangleq {\bf h}_{{\rm e},k}^T\left({\bf b}_{\rm d} + \sum_{i=1}^N{\bf b}_{{\rm n},i}e^{j\left(\phi_{{\rm n},i}-\phi_{\rm d}\right)}\right)$, respectively, ${\tilde y}_{{\rm e},k}, \forall k,$ will lay in the red zone in Fig.~\ref{dest_eve} if either of the following two constraints is satisfied
\begin{subequations}
\begin{align}
\!\!\!\phi_{{\rm e},k} \le -\theta & \Longrightarrow \frac{- \alpha_{{\rm I},k}}{\alpha_{{\rm R},k}-\tilde{\Gamma}_{{\rm e},k}} \le \tan\theta, \forall k, ~ \text{if} ~ \alpha_{{\rm I},k} < 0,\\
\!\!\!\phi_{{\rm e},k} \ge \theta & \Longrightarrow \frac{\alpha_{{\rm I},k}}{\alpha_{{\rm R},k}-\tilde{\Gamma}_{{\rm e},k}} \ge \tan\theta, \forall k, ~ \text{if} ~ \alpha_{{\rm I},k} > 0,
\end{align}
\end{subequations}
That is, the SINR restriction constraints at the Eves can be represented by the following system of inequalities
\begin{subequations}\label{sinr_conste}
\begin{multline}\label{sinr_conste_n}
-\Im \left\{{\bf h}_{{\rm e},k}^T\left({\bf b}_{\rm d} + \sum_{i=1}^N{\bf b}_{{\rm n},i}e^{j\left(\phi_{{\rm n},i}-\phi_{\rm d}\right)}\right)\right\}\\
\le \left(\Re \left\{{\bf h}_{{\rm e},k}^T\left({\bf b}_{\rm d} + \sum_{i=1}^N{\bf b}_{{\rm n},i}e^{j\left(\phi_{{\rm n},i}-\phi_{\rm d}\right)}\right)\right\} - \sigma_{\rm e}\sqrt{\Gamma_{{\rm e},k}}\right)\\
\times\tan\theta, \forall k.
\end{multline}
\begin{multline}\label{sinr_conste_p}
\Im \left\{{\bf h}_{{\rm e},k}^T\left({\bf b}_{\rm d} + \sum_{i=1}^N{\bf b}_{{\rm n},i}e^{j\left(\phi_{{\rm n},i}-\phi_{\rm d}\right)}\right)\right\}\\
\ge \left(\Re \left\{{\bf h}_{{\rm e},k}^T\left({\bf b}_{\rm d} + \sum_{i=1}^N{\bf b}_{{\rm n},i}e^{j\left(\phi_{{\rm n},i}-\phi_{\rm d}\right)}\right)\right\} - \sigma_{\rm e}\sqrt{\Gamma_{{\rm e},k}}\right)\\
\times\tan\theta, \forall k.
\end{multline}
\end{subequations}
where $\tilde{\Gamma}_{{\rm e},k} \triangleq \sigma_{\rm e}\sqrt{\Gamma_{{\rm e},k}}$. Thus exploiting the knowledge of the interfering signals (AN in this case), the constructive AN-based precoding design problem with secrecy power minimization objective can be formulated as
\begin{subequations}\label{minP_dest}
\begin{align}
{\bf{P2:}} ~~ \min_{{\bf b}_{\rm d},\{{\bf b}_{{\rm n},i}\}} ~&~ \left\|{\bf b}_{\rm d} + \sum_{i=1}^N{\bf b}_{{\rm n},i}e^{j\left(\phi_{{\rm n},i}-\phi_{\rm d}\right)} \right\|^2\label{minP_dest_o}
\\
{\rm s.t.} ~&~ \left|\Im \left\{{\bf h}_{\rm d}^T\left({\bf b}_{\rm d} + \sum_{i=1}^N{\bf b}_{{\rm n},i}e^{j\left(\phi_{{\rm n},i}-\phi_{\rm d}\right)}\right)\right\}\right|\nonumber\\
& \le \left(\Re \left\{{\bf h}_{\rm d}^T\left({\bf b}_{\rm d} + \sum_{i=1}^N{\bf b}_{{\rm n},i}e^{j\left(\phi_{{\rm n},i}-\phi_{\rm d}\right)}\right)\right\}\right.\nonumber\\
&\qquad\qquad\qquad \left. - \sigma_{\rm d}\sqrt{\Gamma_{\rm d}}\right)\tan\theta, \label{minP_dest_c1}\\
& \eqref{sinr_conste_n} ~ \text{and}~ \eqref{sinr_conste_p} ~ \text{satisfied}. \label{minP_dest_c2}
\end{align}
\end{subequations}
Problem \eqref{minP_dest} is a standard second-order cone program, thus can be efficiently solved using interior-point based solvers \cite{cvx}. Note that by the inclusion of the data vectors, the secrecy constraints are guaranteed on a symbol-by-symbol basis, as opposed to the conventional statistical secrecy \cite{goel_an, qli_spatial, jrnl_secrecy, jrnl_secrecy_sinr}. Moreover, the virtual multicasting concept introduced in \cite{const_int_kn_int} for a MISO broadcast system is not applicable to the secrecy beamforming problem \eqref{minP_const} or \eqref{minP_dest}.

\section{Robust Constructive-Destructive Interference Precoding}\label{sec_rob_const_dest}
In the previous sections, it was assumed that perfect CSI of all the nodes is available at the transmitter. However, that is a very strict assumption for many practical wireless communication systems. In particular, obtaining the perfect Eves' CSI is always a challenging task. Hence in this section, we study robust AN precoding design for scenarios when the available CSI is imperfect.

We model the imperfect CSI considering the widely used Gaussian channel error model such that the channel error vectors have circularly symmetric complex Gaussian (CSCG) distribution. Thus, the actual channels between the BS and the IR can be modeled as
\begin{equation}
{\bf h}_{\rm d} = \hat{\bf h}_{\rm d} + {\bf e}_{\rm d},
\end{equation}
and that between the BS and the $k$th Eve can be modelled as
\begin{equation}
{\bf h}_{{\rm e},k} = \hat{\bf h}_{{\rm e},k} + {\bf e}_{{\rm e},k}, \forall k,
\end{equation}
where $\hat{\bf h}_{\rm d}$ and $\hat{\bf h}_{{\rm e},k}, \forall k,$ denote the imperfect estimated CSI available at the BS and ${\bf e}_{\rm d}, ~ {\bf e}_{{\rm e},k} \in\mathbb{C}^{N_{\rm T}\times 1}, \forall k$, represent the channel uncertainties such that $\|{\bf e}_{\rm d}\|^2 \le \varepsilon_{\rm d}^2$, and $\|{\bf e}_{{\rm e},k}\|^2 \le \varepsilon_{{\rm e}}^2, \forall k,$ respectively.

\subsection{Conventional AN-Aided Robust Secure Precoding}
Conventional AN-aided downlink robust secrecy power minimization problem with SINR constraints is formulated as
\begin{subequations}\label{rob_minP_conv}
\begin{eqnarray}
\min_{{\bf b}_{\rm d},\{{\bf b}_{{\rm n},i}\}} \!\!\!& &\!\!\! \left\|{\bf b}_{\rm d}\right\|^2 + \sum_{i=1}^N\left\|{\bf b}_{{\rm n},i}\right\|^2\label{rob_minP_conv_o}\\
{\rm s.t.} \!\!\!& &\!\!\! \min_{\|{\bf e}_{\rm d}\| \le \varepsilon_{\rm d}}\frac{\left|{\bf h}_{\rm d}^T{\bf b}_{\rm d}\right|^2}{\sum_{i=1}^N\left|{\bf h}_{\rm d}^T{\bf b}_{{\rm n},i}\right|^2 + \sigma_{\rm d}^2}\ge \Gamma_{\rm d},\label{rob_minP_conv_c1}\\
\!\!\!& &\!\!\! \max_{\|{\bf e}_{{\rm e},k}\| \le \varepsilon_{{\rm e}}}\frac{\left|{\bf h}_{{\rm e},k}^T{\bf b}_{\rm d}\right|^2}{\sum_{i=1}^N\left|{\bf h}_{{\rm e},k}^T{\bf b}_{{\rm n},i}\right|^2 + \sigma_{\rm e}^2} \le \Gamma_{{\rm e},k}.\label{rob_minP_conv_c2}
\end{eqnarray}
\end{subequations}
Due to the spherical channel uncertainty model, constraints \eqref{rob_minP_conv_c1} and \eqref{rob_minP_conv_c2} actually involve infinitely many constraints which makes the problem \eqref{rob_minP_conv} very difficult to solve. However, applying $\mathcal{S}$-procedure \cite[Lemma~2]{jrnl_secrecy}, the inequality constraints in \eqref{rob_minP_conv} can be transformed into convex linear matrix inequality constraints and thus problem \eqref{rob_minP_conv} can be readily solved using existing solvers. It has been proved in \cite{jrnl_secrecy_sinr} that whenever problem \eqref{rob_minP_conv} is feasible, the corresponding transmit precoding solution is of rank-one hence optimal.

\subsection{Constructive AN-Aided Robust Secure Precoding}
In this section, we aim at robust precoding design such that the AN is constructive to the IR while destructive to the Eves with imperfect knowledge of all CSI. With the deterministic channel uncertainty model described above, we consider worst-case based robust design. Thus the constructive AN based robust power minimization problem can be formulated as given in equation \eqref{rob_minP_dest}.

\begin{figure*}[tb]
\begin{subequations}\label{rob_minP_dest}
\begin{align}
\min_{{\bf b}_{\rm d},\{{\bf b}_{{\rm n},i}\}} ~&~ \left\|{\bf b}_{\rm d} + \sum_{i=1}^N{\bf b}_{{\rm n},i}e^{j\left(\phi_{{\rm n},i}-\phi_{\rm d}\right)} \right\|^2\label{rob_minP_dest_o}
\\
{\rm s.t.} ~&~ \left|\Im \left\{{\bf h}_{\rm d}^T\left({\bf b}_{\rm d} + \sum_{i=1}^N{\bf b}_{{\rm n},i}e^{j\left(\phi_{{\rm n},i}-\phi_{\rm d}\right)}\right)\right\}\right| \le \left(\Re \left\{{\bf h}_{\rm d}^T\left({\bf b}_{\rm d} + \sum_{i=1}^N{\bf b}_{{\rm n},i}e^{j\left(\phi_{{\rm n},i}-\phi_{\rm d}\right)}\right)\right\} - \sigma_{\rm d}\sqrt{\Gamma_{\rm d}}\right)\tan\theta, \nonumber\\
& \qquad\qquad\qquad \forall {\|{\bf e}_{\rm d}\| \le \varepsilon_{\rm d}}, \label{rob_minP_dest_c1}\\
& -\Im \left\{{\bf h}_{{\rm e},k}^T\left({\bf b}_{\rm d} + \sum_{i=1}^N{\bf b}_{{\rm n},i}e^{j\left(\phi_{{\rm n},i}-\phi_{\rm d}\right)}\right)\right\} \le \left(\Re \left\{{\bf h}_{{\rm e},k}^T\left({\bf b}_{\rm d} + \sum_{i=1}^N{\bf b}_{{\rm n},i}e^{j\left(\phi_{{\rm n},i}-\phi_{\rm d}\right)}\right)\right\} - \sigma_{\rm e}\sqrt{\Gamma_{{\rm e},k}}\right)\tan\theta, \nonumber\\
& \qquad\qquad\qquad\forall {\|{\bf e}_{{\rm e},k}\| \le \varepsilon_{{\rm e}}}, \forall k, \label{rob_minP_dest_c2}\\
& \Im \left\{{\bf h}_{{\rm e},k}^T\left({\bf b}_{\rm d} + \sum_{i=1}^N{\bf b}_{{\rm n},i}e^{j\left(\phi_{{\rm n},i}-\phi_{\rm d}\right)}\right)\right\} \ge \left(\Re \left\{{\bf h}_{{\rm e},k}^T\left({\bf b}_{\rm d} + \sum_{i=1}^N{\bf b}_{{\rm n},i}e^{j\left(\phi_{{\rm n},i}-\phi_{\rm d}\right)}\right)\right\} - \sigma_{\rm e}\sqrt{\Gamma_{{\rm e},k}}\right)\tan\theta, \nonumber\\
& \qquad\qquad\qquad \hfill \forall {\|{\bf e}_{{\rm e},k}\| \le \varepsilon_{{\rm e}}}, \forall k. \label{rob_minP_dest_c3}
\end{align}
\end{subequations}
\hrulefill \normalsize
\end{figure*}
Note that the information and the AN beamforming vectors appear in identical form in the objective functions as well as in the constraints in problem \eqref{rob_minP_dest}. Denoting ${\bf b} \triangleq {\bf b}_{\rm d} + \sum_{i=1}^N{\bf b}_{{\rm n},i}e^{j\left(\phi_{{\rm n},i}-\phi_{\rm d}\right)}$, the problem can thus be represented as
\begin{subequations}\label{rob_minP_dest2}
\begin{align}
\min_{{\bf b}_{\rm d},\{{\bf b}_{{\rm n},i}\}} ~&~ \left\|{\bf b}\right\|^2\label{rob_minP_dest2_o}
\\
{\rm s.t.} ~&~ \left|\Im \left\{\left(\hat{\bf h}_{\rm d} + {\bf e}_{\rm d}\right)^T{\bf b}\right\}\right| \le \left(\Re \left\{\left(\hat{\bf h}_{\rm d} + {\bf e}_{\rm d}\right)^T{\bf b}\right\}\right.\nonumber\\
& \left. - \sigma_{\rm d}\sqrt{\Gamma_{\rm d}}\right) \tan\theta, \quad \forall {\|{\bf e}_{\rm d}\| \le \varepsilon_{\rm d}}, \label{rob_minP_dest2_c1}\\
& -\Im \left\{\left(\hat{\bf h}_{{\rm e},k} + {\bf e}_{{\rm e},k}\right)^T{\bf b}\right\} \le \left(\Re \left\{\left(\hat{\bf h}_{{\rm e},k} + {\bf e}_{{\rm e},k}\right)^T\right.\right.\nonumber\\
& \left.\left.\times {\bf b}\right\} - \sigma_{\rm e}\sqrt{\Gamma_{{\rm e},k}}\right) \tan\theta, ~ \forall {\|{\bf e}_{{\rm e},k}\| \le \varepsilon_{{\rm e}}}, \forall k, \label{rob_minP_dest2_c2}\\
& \Im \left\{\left(\hat{\bf h}_{{\rm e},k} + {\bf e}_{{\rm e},k}\right)^T{\bf b}\right\} \ge \left(\Re \left\{\left(\hat{\bf h}_{{\rm e},k} + {\bf e}_{{\rm e},k}\right)^T{\bf b}\right\}\right.\nonumber\\
& \left.  - \sigma_{\rm e}\sqrt{\Gamma_{{\rm e},k}}\right) \tan\theta, \quad \forall {\|{\bf e}_{{\rm e},k}\| \le \varepsilon_{{\rm e}}}, \forall k. \label{rob_minP_dest2_c3}
\end{align}
\end{subequations}
Considering the real and imaginary parts of each complex vector separately, we have
\begin{align}
{\bf h}_{\rm d} & = \hat{\bf h}_{R,{\rm d}} + j\hat{\bf h}_{I,{\rm d}} + {\bf e}_{R,{\rm d}} + j{\bf e}_{I,{\rm d}},\\
{\bf b} & = {\bf b}_R + j{\bf b}_I,
\end{align}
where the subscripts $R$ and $I$ indicate the real and imaginary components of the corresponding complex notation, respectively. As such, we have,
\begin{align}
\Re \left\{\left(\hat{\bf h}_{\rm d} + {\bf e}_{\rm d}\right)^T{\bf b}\right\} & = \hat{\bf h}_{R,{\rm d}}^T{\bf b}_R - \hat{\bf h}_{I,{\rm d}}^T{\bf b}_I + {\bf e}_{R,{\rm d}}^T{\bf b}_R - {\bf e}_{I,{\rm d}}^T{\bf b}_I \nonumber\\
& = \tilde{\bf h}_{\rm d}^T{\bf b}_1 + \tilde{\bf e}_{\rm d}^T{\bf b}_1,
\end{align}
where $\tilde{\bf h}_{\rm d} \triangleq \left[\hat{\bf h}_{R,{\rm d}}^T ~~ \hat{\bf h}_{I,{\rm d}}^T\right]^T$, $\tilde{\bf e}_{\rm d} \triangleq \left[{\bf e}_{R,{\rm d}}^T ~~ {\bf e}_{I,{\rm d}}^T\right]^T$, and ${\bf b}_1 \triangleq \left[{\bf b}_R^T ~~ -{\bf b}_I^T\right]^T$. Similarly,
\begin{align}
\Im \left\{\left(\hat{\bf h}_{\rm d} + {\bf e}_{\rm d}\right)^T{\bf b}\right\} & = \hat{\bf h}_{R,{\rm d}}^T{\bf b}_R + \hat{\bf h}_{I,{\rm d}}^T{\bf b}_I + {\bf e}_{R,{\rm d}}^T{\bf b}_R + {\bf e}_{I,{\rm d}}^T{\bf b}_I \nonumber\\
& = \tilde{\bf h}_{\rm d}^T{\bf b}_2 + \tilde{\bf e}_{\rm d}^T{\bf b}_2,
\end{align}
with ${\bf b}_2 \triangleq \left[{\bf b}_R^T ~~ {\bf b}_I^T\right]^T$. Thus the constraint \eqref{rob_minP_dest2_c1} can be explicitly expressed as the following two constraints
\begin{align}
\max_{\|{\bf e}_{\rm d}\| \le \varepsilon_{\rm d}} \tilde{\bf h}_{\rm d}^T{\bf b}_2 + \tilde{\bf e}_{\rm d}^T{\bf b}_2 - \left(\tilde{\bf h}_{\rm d}^T{\bf b}_1 + \tilde{\bf e}_{\rm d}^T{\bf b}_1\right) \tan\theta \nonumber\\
+ \sigma_{\rm d}\sqrt{\Gamma_{\rm d}} \le 0\\
\max_{\|{\bf e}_{\rm d}\| \le \varepsilon_{\rm d}} -\tilde{\bf h}_{\rm d}^T{\bf b}_2 - \tilde{\bf e}_{\rm d}^T{\bf b}_2 - \left(\tilde{\bf h}_{\rm d}^T{\bf b}_1 + \tilde{\bf e}_{\rm d}^T{\bf b}_1\right) \tan\theta \nonumber\\
+ \sigma_{\rm d}\sqrt{\Gamma_{\rm d}} \le 0.
\end{align}
Similarly, the constraints \eqref{rob_minP_dest2_c2} and \eqref{rob_minP_dest2_c3} can be, respectively, rewritten as
\begin{align}
\max_{\|{\bf e}_{{\rm e},k}\| \le \varepsilon_{\rm e}} -\tilde{\bf h}_{{\rm e},k}^T{\bf b}_2 - \tilde{\bf e}_{{\rm e},k}^T{\bf b}_2 - & \left(\tilde{\bf h}_{{\rm e},k}^T{\bf b}_1 + \tilde{\bf e}_{{\rm e},k}^T{\bf b}_1\right) \tan\theta \nonumber\\
& + \sigma_{\rm e}\sqrt{\Gamma_{{\rm e},k}} \le 0\\
\min_{\|{\bf e}_{{\rm e},k}\| \le \varepsilon_{\rm e}} \tilde{\bf h}_{{\rm e},k}^T{\bf b}_2 + \tilde{\bf e}_{{\rm e},k}^T{\bf b}_2 - & \left(\tilde{\bf h}_{{\rm e},k}^T{\bf b}_1 + \tilde{\bf e}_{{\rm e},k}^T{\bf b}_1\right) \tan\theta \nonumber\\
& + \sigma_{\rm e}\sqrt{\Gamma_{{\rm e},k}} \ge 0.
\end{align}
By replacing the CSI error bounds in these constraints, the robust problem \eqref{rob_minP_dest2} can be reformulated as
\begin{subequations}\label{rob_minP_dest3}
\begin{align}
\min_{{\bf b}_{1},{\bf b}_{2}} ~&~ \left\|{\bf b}_2\right\|^2 ~~ {\rm s.t.}\label{rob_minP_dest3_o}
\\
& \tilde{\bf h}_{\rm d}^T{\bf b}_2 - \tilde{\bf h}_{\rm d}^T{\bf b}_1\tan\theta  + \varepsilon_{\rm d}\left\|{\bf b}_2 - {\bf b}_1 \tan\theta\right\| \nonumber\\
& + \sigma_{\rm d}\sqrt{\Gamma_{\rm d}} \le 0\label{rob_minP_dest3_c0}\\
& - \tilde{\bf h}_{\rm d}^T{\bf b}_2 - \tilde{\bf h}_{\rm d}^T{\bf b}_1\tan\theta  + \varepsilon_{\rm d}\left\|{\bf b}_2 + {\bf b}_1 \tan\theta\right\| \nonumber\\
& + \sigma_{\rm d}\sqrt{\Gamma_{\rm d}} \le 0, \label{rob_minP_dest3_c1}\\
& -\tilde{\bf h}_{{\rm e},k}^T{\bf b}_2 - \tilde{\bf h}_{{\rm e},k}^T{\bf b}_1\tan\theta - \varepsilon_{\rm e}\left\|{\bf b}_2 + {\bf b}_1\right\| \tan\theta \nonumber\\
& + \sigma_{\rm e}\sqrt{\Gamma_{{\rm e},k}} \le 0\label{rob_minP_dest3_c2}\\
& \tilde{\bf h}_{{\rm e},k}^T{\bf b}_2 + \tilde{\bf h}_{{\rm e},k}^T{\bf b}_1\tan\theta - \varepsilon_{\rm e}\left\|{\bf b}_2 + {\bf b}_1\right\| \tan\theta \nonumber\\
& + \sigma_{\rm e}\sqrt{\Gamma_{{\rm e},k}} \ge 0. \label{rob_minP_dest3_c3}
\end{align}
\end{subequations}
The SOCP problem \eqref{rob_minP_dest3} can be efficiently solved using existing solvers \cite{cvx}.

\section{Simulation Results}\label{sec_sim}
This section presents numerical simulation results to evaluate the performance of the proposed constructive interference based PLS algorithms in a MISO wiretap channel. For comparison, conventional secure precoding performances have also been included. For simplicity, it was assumed that $\Gamma_{{\rm e},k} = \Gamma_{\rm e}, ~\forall k$ and $\sigma_{\rm d}^2 = \sigma_{\rm e}^2 = 1$. Unless otherwise specified, $N=3$ and QPSK is the modulation scheme considered. All the estimated channel vectors are generated as independent and identically distributed complex Gaussian random variables with mean zero and the TGn path-loss model for urban cellular environment is adopted considering a path-loss exponent of $2.7$ \cite{ch_model}.
All simulation results are averaged over $1000$ independent channel realizations, unless explicitly mentioned. In the following simulations, we compare the performance of the proposed approaches with that of the conventional AN-aided precoding scheme in \cite{qli_qos} as the benchmark. Specifically, we denote the conventional precoding schemes as `Conv Prec', the constructive interference based precoding scheme developed in Section~\ref{sec_prob_const_int} as `Const Prec', and the destructive interference based scheme in Section~\ref{sec_dest_int} as `Const-Dest Prec' in the figures below.

\begin{figure}
\centering
\includegraphics*[width=8cm]{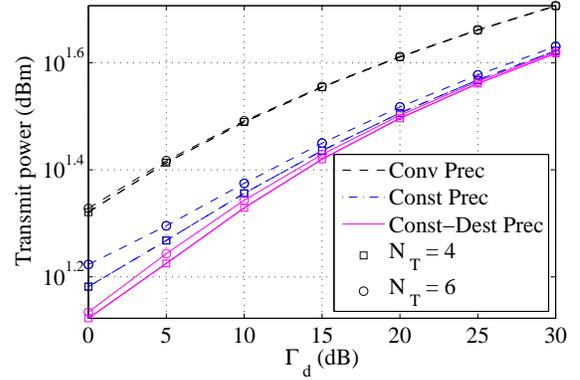}
\caption{Transmit power $P_{\rm T}$ versus required SINR at IR $\Gamma_{\rm d}$ with $N_{\rm T} = 8, K = 4, 6$, and $\Gamma_{\rm e} = 5$ (dB).}\label{fig_Pt_Gmm}
\end{figure}

We start the performance analysis of the proposed constructive interference based secure AN precoding schemes assuming perfect CSI of all the nodes available at the transmitter.  Fig.~\ref{fig_Pt_Gmm} shows the average transmit power versus the SINR requirement at the IR required for the proposed constructive-only (in problem \eqref{minP_const}) as well as the constructive-destructive AN precoding optimization scheme (in problem \eqref{minP_dest}) as compared with the conventional AN precoding scheme (in problem \eqref{minP_conv2}) for $K = 4$ and $6$. Other parameters are set as $N_{\rm T} = 8$ and $\Gamma_{\rm e} = 5$ (dB). It can be observed that the proposed constructive interference algorithms achieve significant power gains compared to the conventional AN precoding scheme. Also, if there exist more eavesdroppers, additional power is needed in all optimization schemes to block the interception.

\begin{figure}
\centering
\includegraphics*[width=8cm]{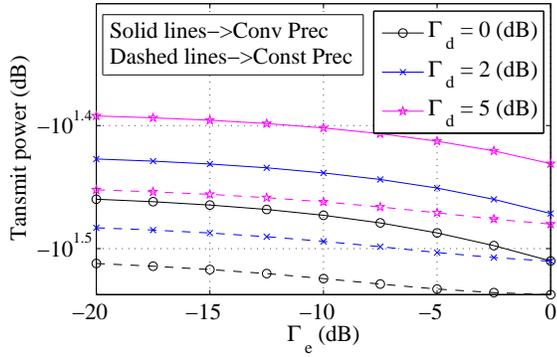}
\caption{Transmit power $P_{\rm T}$ versus required SINR $\Gamma_{\rm e}$ with $N_{\rm T} = 6, K = 4$, and $\Gamma_{\rm d} = 20$ (dB).}\label{fig_Pt_Gme}
\end{figure}

In the next example, we examine the transmit power requirement against the maximum allowable eavesdropping SINR $\Gamma_{\rm e}$. Fig.~\ref{fig_Pt_Gme} plots the average transmit power $P_{\rm T}$ versus $\Gamma_{\rm e}$ for $N_{\rm T} = 6$, $K = 4$ and different values of $\Gamma_{\rm d}$. The results in Fig.~\ref{fig_Pt_Gme}  are consistent with those in Fig.~\ref{fig_Pt_Gmm} in the sense that increased SINR threshold at the IR requires higher transmit power. However, with the increase in the allowable SINR threshold at the Eves, the required transmit power gradually decreases due to the relaxed eavesdropping constraints. Also, in any case, the constructive interference based precoding schemes outperform the conventional AN-aided secure precoding schemes.

\begin{figure}
\centering
\includegraphics*[width=8cm]{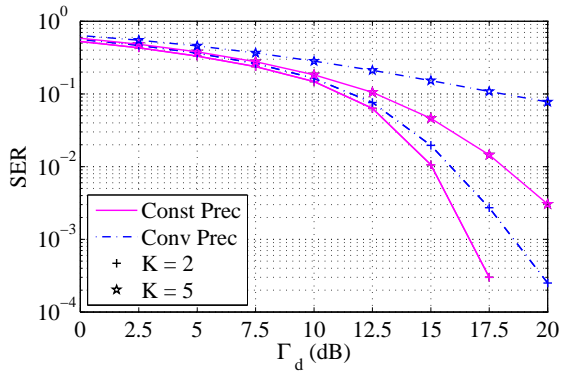}
\caption{SER versus SINR threshold $\Gamma_{\rm d}$ with $N_{\rm T} = 6, K = 2, 5$, and $\Gamma_{\rm e} = 5$ (dB).}\label{fig_ser_Gmm}
\end{figure}

Next, we analyze SER performance of the proposed scheme as opposed to the conventional precoding schemes. The results in Fig.~\ref{fig_ser_Gmm} indicate that the SER in all the schemes decreases in line with the SINR threshold at the IR. Notably, at a SER of $10^{-3}$, the constructive interference based secure precoding scheme achieves almost $2.4$ dB gain compared to the conventional approach. Also, it is no surprise that a larger number of Eves ($K$) has a negative impact on the SER since we need to employ more resources to block a larger set of Eves and there will be more non-active constraints for the eavesdropping SINR.

\begin{figure}
\centering
\includegraphics*[width=8cm]{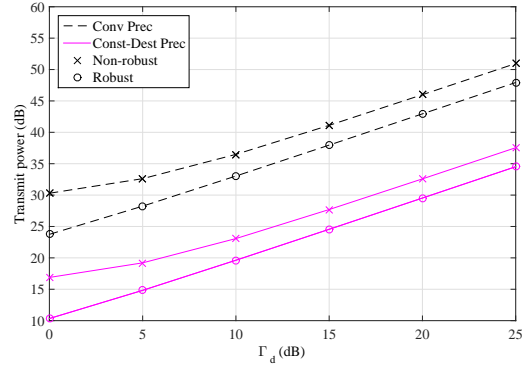}
\caption{Transmit power $P_{\rm T}$ versus required SINR $\Gamma_{\rm d}$ with $N_{\rm T} = 6, K = 3$, $\Gamma_{\rm e} = 5$ (dB), and $\varepsilon_{\rm d} = 0.1, \varepsilon_{\rm e} = 0.3$.}\label{fig_Pt_Gmm_Rob}
\end{figure}

Finally, we analyze the performance of the proposed robust beamforming design with $N_{\rm T} = 6, K = 3$, $\Gamma_{\rm e} = 5$ (dB), and $\varepsilon_{\rm d} = 0.1, \varepsilon_{\rm e} = 0.3$, when imperfect CSI is available at the BS. In Fig.~\ref{fig_Pt_Gmm_Rob}, the robust schemes indicate the solution to the problems \eqref{rob_minP_conv} and \eqref{rob_minP_dest3}, respectively, for conventional and constructive AN based precoding schemes. On the other hand, the `Non-robust' scheme is designed treating the imperfect channel estimates available at the BS as the perfect CSI, hence yields noticeable performance degradation. However, the proposed constructive interference based robust secure beamforming schemes demonstrate significant transmit power gain.

\section{Conclusions}\label{sec_con}
We proposed the novel idea of designing the AN-aided secure precoding schemes as constructive to the IR and destructive to the Eves at the same time. This introduces a major breakthrough in the conventional approach of transmitting AN for improving PLS. The concept opens up new opportunities for expanding the secrecy rate regions. We studied the downlink transmit power minimization problem considering both perfect and imperfect CSI at the BS. Simulation results demonstrated that significant performance gain is achievable by the proposed constructive AN precoding schemes compared to the conventional schemes and have established the proposed approach as a new dimension in the design of PLS.

\bibliographystyle{IEEEtran}\footnotesize{
\bibliography{IEEEabrv,refdb}}%

\end{document}